\documentclass[twoside,a4paper,11pt]{torus2012}
\usepackage{graphicx}
\usepackage[colorlinks=true,citecolor=blue]{hyperref}
\usepackage{movie15}
\begin{document}
\pagenumbering{arabic}
\pagestyle{myheadings}
\thispagestyle{empty}
{\flushleft\includegraphics[width=\textwidth,bb=58 650 590 680]{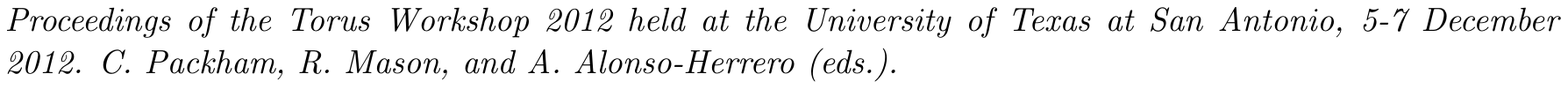}}
\vspace*{0.2cm}
\begin{flushleft}
{\bf {\LARGE
%
Mid-$J$ CO Emission in Nearby Seyfert Galaxies
%
}\\
\vspace*{1cm}
%

Miguel Pereira-Santaella$^{1}$, 
Luigi Spinoglio$^{1}$, 
Gemma Busquet$^{1}$,
Jason Glenn$^{2}$, 
Kate Isaak$^{3}$,
Julia Kamenetzky$^{2}$,
Naseem Rangwala$^{2}$,
Maximilien~R.~P. Schirm$^{4}$,
Maarten Baes$^{5}$, 
Michael~J. Barlow$^{6}$, 
Alessandro Boselli$^{7}$, 
Asantha Cooray$^{8}$, 
and
Diane Cormier$^{9}$
%
}\\
\vspace*{0.5cm}
%
\small
$^{1}$
Istituto di Astrofisica e Planetologia Spaziali, INAF, Via Fosso del Cavaliere 100, I-00133 Roma, Italy. e-mail: miguel.pereira@ifsi-roma.inaf.it \\
$^{2}$
Center for Astrophysics and Space Astronomy, 389-UCB, University of Colorado, Boulder, CO 80303, USA \\
$^{3}$
ESA Astrophysics Missions Division, ESTEC, PO Box 299, 2200 AG Noordwijk, The Netherlands \\
$^{4}$
Dept. of Physics \& Astronomy, McMaster University, Hamilton, Ontario, L8S 4M1, Canada \\
$^{5}$
Sterrenkundig Observatorium, Universiteit Gent, Krijgslaan 281 S9, B-9000 Gent, Belgium \\
$^{6}$
Department of Physics and Astronomy, University College London, Gower Street, London WC1E 6BT,UK \\
$^{7}$
Laboratoire d'Astrophysique de Marseille - LAM, Universit\`e d'Aix-Marseille \& CNRS, UMR7326, 38 rue F. Joliot-Curie, 13388, Marseille Cedex 13, France \\
$^{8}$
Center for Cosmology, Department of Physics and Astronomy, University of California, Irvine, CA 92697, USA \\
$^{9}$
Laboratoire AIM, CEA/DSM-CNRS-Universit\`e Paris Diderot, Irfu/Service d'Astrophysique, CEA Saclay, F-91191, Gif-sur-Yvette, France \\
%
\end{flushleft}
%
\markboth{
Mid-$J$ CO Emission in Nearby Seyfert Galaxies
}{ 
%
Pereira-Santaella et al.
%
}
\thispagestyle{empty}
\vspace*{0.4cm}
\begin{minipage}[l]{0.09\textwidth}
\ 
\end{minipage}
\begin{minipage}[r]{0.9\textwidth}
\vspace{1cm}
\section*{Abstract}{\small
We study for the first time the complete sub-millimeter spectra (450 GHz to 1550 GHz) of a sample of nearby
active galaxies observed with the SPIRE Fourier Transform Spectrometer (SPIRE\slash FTS) onboard
\textit{Herschel}\footnote{\textit{Herschel} is an ESA space observatory with science instruments provided by European-led Principal Investigator consortia and with important participation from NASA.}.
The CO ladder (from $J_{\rm up}$ = 4 to 12) is the most prominent spectral feature in this range. 
These CO lines probe warm molecular gas that can be heated by ultraviolet photons, shocks, or X-rays originated in the active galactic nucleus or in young star-forming regions.
In these proceedings we investigate the physical origin of the CO emission using the averaged CO spectral line energy distribution (SLED) of six Seyfert galaxies. We use a radiative transfer model assuming an isothermal homogeneous medium to estimate the molecular gas conditions. We also compare this CO SLED with the predictions of photon and X-ray dominated region (PDR and XDR) models.
%
\normalsize}
\end{minipage}
%
%
%
\section{Introduction}
To date, the sub-millimeter (sub-mm) spectrum of galaxies has not been fully explored, in part due to the low transparency of the atmosphere in this range.
The SPIRE Fourier transform spectrometer (SPIRE\slash FTS) \cite{Griffin2010SPIRE,Naylor2010,Swinyard2010} onboard the \textit{Herschel} Space Observatory \cite{Pilbratt2010Herschel} fills this gap, covering the spectral range between 450 and 1440 GHz (210 and 670 $\mu$m).

In the SPIRE\slash FTS spectra the most obvious spectral feature is the mid-$J$ CO ladder ($J_{\rm up}$ = 4 to 13). CO is one of the most abundant molecules in the interstellar medium; therefore it is considered a good tracer of the molecular gas phase. Specifically, the mid-$J$ CO line emission originate from warm molecular gas (upper-level energies ranging from 55 to 500\,K above the ground state) that can be excited by ultraviolet (UV) photons in photon dominated regions (PDRs; e.g., \cite{Wolfire2010}), X-rays in X-ray dominated regions (XDRs; e.g., \cite{Meijerink2006}) or by shocks (e.g., \cite{Flower2010}). 

In these proceedings we analyze the average CO spectral line energy distribution (SLED) of local Seyfert galaxies, including both low- and mid-$J$ CO transitions, in order to understand the physical conditions of the warm molecular gas and the dominant excitation mechanism in this class of galaxies.

\section{Sub-mm Spectra of Local Seyfert Galaxies}

We obtained sub-mm SPIRE\slash FTS spectra of eleven nearby ($d=12-160$\,Mpc) Seyfert galaxies selected from the 12\,$\mu$m Galaxy Sample \cite{Rush1993} through two guaranteed time projects: \textit{``Bright Seyfert Nuclei: FTS spectroscopy''} (PI: L. Spinoglio) and \textit{``Physical Processes in the Interstellar Medium of Very Nearby Galaxies''} (PI: C.~D. Wilson).
Our sample includes galaxies with different nuclear activity classes in similar proportions. Their infrared (IR) luminosities\footnote{The IR luminosity refers to the 8--1000\,$\mu$m galaxy integrated luminosity while the hard X-ray luminosity is the 2--10\,keV luminosity.\label{foot_lum}} range from 10$^{10}$ to 10$^{12}$\,$L_{\rm \odot}$ and their hard X-ray luminosities$^{\ref{foot_lum}}$ from 10$^{40}$ to 10$^{43}$\,erg\,s$^{-1}$. The SPIRE\slash FTS beam size is 20--40$^{\prime\prime}$ that corresponds to 1.4--32\,kpc at the distance of our galaxies. More details on the sample and data reduction are given in \cite{Pereira2013}.

In the SPIRE\slash FTS spectra of these galaxies we detected transitions of several molecules. The brightest lines are those produced by CO, but lines from other molecules like H$_2$O, OH$^{+}$, and HF are often detected. In addition, atomic lines from neutral gas ([C~{\small I}] at 492 and 809\,GHz) and ionized gas ([N~{\small II}] at 1461\,GHz) are bright in these galaxies \cite{Pereira2013}. In these proceedings we focus on the mid-$J$ CO emission.

\begin{figure}
\center
\includegraphics[width=13.cm]{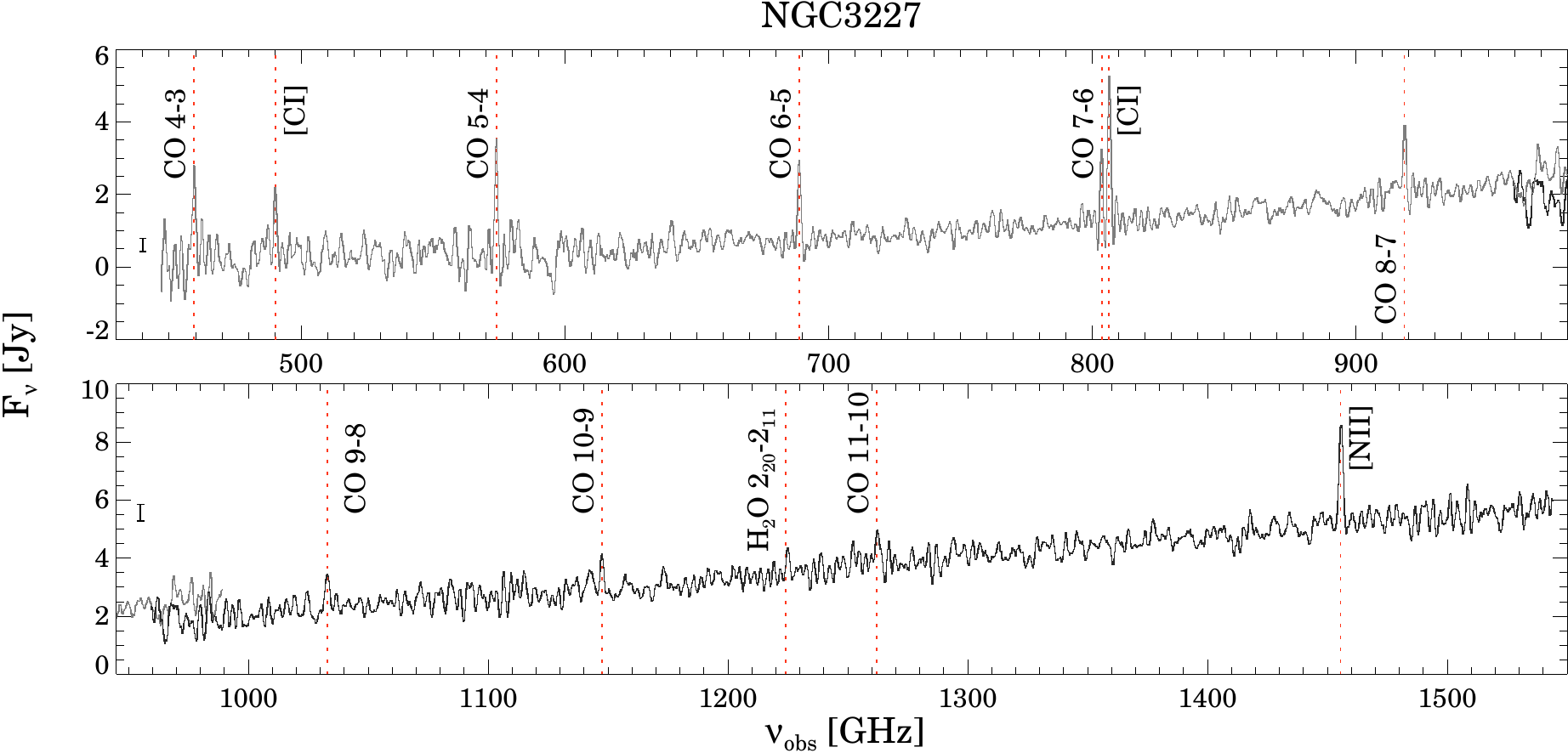}
\caption{\label{fig1} Observed SPIRE\slash FTS spectra of the Seyfert 1.5 NGC~3227. The dashed red lines mark the position of the detected lines. The error bars indicate the median 1$\sigma$ uncertainty. }
\end{figure}

For six of these galaxies we detected more than three CO lines at $>$3$\sigma$ level in their SPIRE\slash FTS spectra (see Figure~\ref{fig1}) that were used, together with ground observations of the low-$J$ CO lines ($J_{\rm up}=1$ to 3), to construct the CO SLEDs for each galaxy. The CO SLEDs of five of them (UGC~05101, NGC~3227, NGC~4388, NGC~7130, and NGC~7582) are remarkably similar. Therefore we averaged them (normalized by their CO luminosity) to construct an average SLED which we used to study the typical conditions in the warm molecular gas in Seyfert galaxies.
We note that the Seyfert type, the AGN luminosity, or the total IR luminosity (mainly related to their star-formation rate) do not seem to affect the CO SLED shape in our sample of galaxies.

\section{CO Emission Modeling}

\subsection{Radiative Transfer Model}
We used the non-LTE radiative transfer code RADEX, with the assumption that the mid-$J$ CO emission originates from a homogeneous medium, to infer the physical conditions of the emitting regions. 
We produced a grid of models that was used to model the average CO SLED. The best fit RADEX model is shown in Figure \ref{fig2} (dotted red line). The derived kinetic temperature is 500$^{+500}_{-250}$\,K and the H$_2$ density 10$^{3.7\pm 1.0}$\,cm$^{-3}$. As expected, these values agree with those obtained from the individual galaxies (see \cite{Pereira2013}). Moreover, these conditions are similar to those observed in the starburst M~82 \cite{Kamenetzky2012}. Differently, in the ultra-luminous IR galaxy Arp~220 the gas temperature is higher ($\sim1300$\,K) \cite{Rangwala2011}, whereas in the Seyfert 2 NGC~1068 it is lower ($\sim90$\,K) \cite{Spinoglio2012}. This is consistent with the mid-$J$ CO emission from our sub-sample of Seyfert galaxies being driven by star-formation activity rather than by the AGN.

\subsection{PDR Models}\label{s:pdr}

\begin{figure}
\center
\includegraphics[width=8.cm]{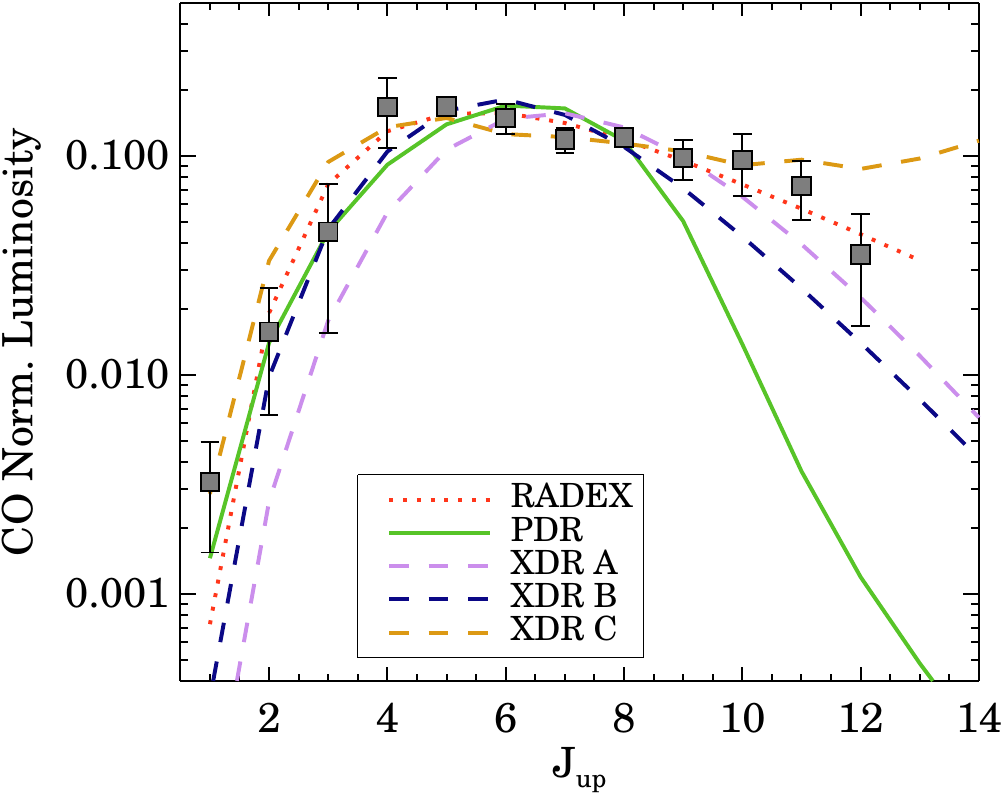}
\caption{\label{fig2} Average CO SLED of our sample of nearby Seyfert galaxies (gray squares) normalized by the total CO luminosity. The dotted red line represents the best-fit RADEX model. The solid green line is the best-fit PDR model \cite{Wolfire2010}. XDR models of \cite{Meijerink2006} are plotted with purple, blue, and orange dashed lines (see Section \ref{s:xdr}).
}
\end{figure}

In star-forming regions where UV photons from young stars heat the ISM we expect PDR excitation of the molecular gas. Therefore we compared the average CO SLED with the PDR models of \cite{Wolfire2010}. In Figure \ref{fig2} we plot the best-fit model (solid green line). The plotted model corresponds to $n_{\rm H_2}= 10^{4.25}$\,cm$^{-3}$ and far-UV flux 10$^{5.5}$\,$G_{0}$, where $G_{\rm 0}=1.6 \times 10^{-3}$\,erg\,cm$^{-2}$\,s$^{-1}$. However, models with $n_{\rm H_2}= 10^{4.2}-10^{5.0}$\,cm$^{-3}$ and $F_{\rm FUV}=10-10^{6.5}\,G_{\rm 0}$\footnote{$F_{\rm FUV}=10^{6.5}$\,$G_{\rm 0}$ is the upper limit of the grid of PDR models, so higher $F_{\rm FUV}$ could be compatible as well.} also provide equally good fits. 
These PDR models provide a good fit to the emission of CO lines with $J_{\rm up} < 8$, however emission from higher-$J$ transitions is clearly underpredicted.

\begin{figure}
\center
\includegraphics[width=8cm]{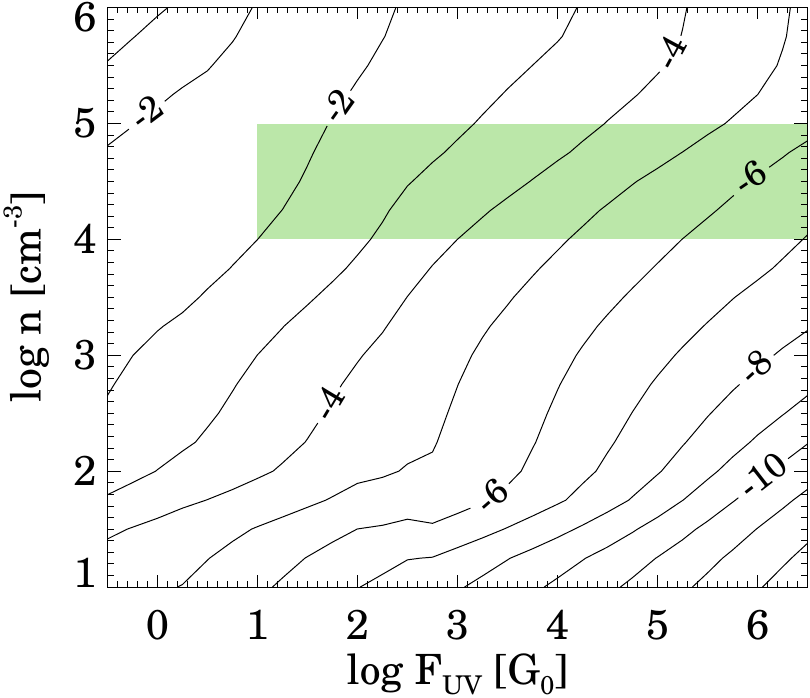}
\caption{\label{fig3} The black contours represent the logarithm of the $L_{\rm CO}$\slash $L_{\rm IR}$ ratio as a function of the incident far-UV flux and gas density. The models capable of fitting the observed CO SLED (up to $J_{\rm up} = 7$) are marked with a shaded green area. We assumed $L_{\rm IR}=2\times L_{\rm FUV}$~\cite{Kaufman1999}.}
\end{figure}

Figure \ref{fig3} shows the $L_{\rm CO}$\slash $L_{\rm IR}$\footnote{$L_{\rm CO}$ is defined as the integrated luminosity of the CO lines from $J_{\rm up} = 4$ to $J_{\rm up} = 12$.\label{foor_lco}} ratio as a function of the incident far-UV flux and gas density. In our galaxies this ratio is $\sim$10$^{-4}$, however this is a lower limit because not all the IR luminosity have to be associated with the warm molecular gas emitting the mid-$J$ CO lines. Consequently, from this figure we can constrain the far-UV radiation field to be between 10 and 10$^{4}$\,$G_{\rm 0}$.

\subsection{XDR Models}\label{s:xdr}

We also fitted our averaged CO SLED using the XDR models of \cite{Meijerink2006}. These authors model three possible scenarios for XDR emission: high-density ($n_{\rm H_2}\sim10^4-10^{6.5}$\,cm$^{-3}$) compact molecular clouds close to the AGN (XDR A); ``traditional'' mid-range density ($n_{\rm H_2}\sim10^3-10^4$\,cm$^{-3}$) molecular clouds (XDR B); and diffuse molecular gas ($n_{\rm H_2}\sim10^2-10^3$\,cm$^{-3}$) where XDR B type clouds are usually embedded (XDR C). The best-fit models are plotted in Figure \ref{fig2} with dashed lines.
The XDR models better reproduce the CO emission for $J_{\rm up} > 8$ than do the PDR models. The best one is the XDR B, although it systematically underpredicts the CO luminosities for $J_{\rm up} > 8$. Conversely, the XDR A model underpredict CO lines with $J_{\rm up} < 6$, but agrees with the observed higher-$J$ emission. 

The $L_{\rm CO}$\slash $L_{\rm X}$\footnote{$L_{\rm X}$ is the 1--100\,keV X-ray luminosity.} ratio, for these two XDR models is $\sim$10$^{-4}$; therefore for our Seyfert galaxies the observed $L_{\rm CO}$ implies X-ray luminosities $100-10\,000$ times higher than those observed. Other XDR models in these grids have higher $L_{\rm CO}$\slash $L_{\rm X}$ ratios (up to $\sim$10$^{-2}$) that would be compatible with the AGN luminosities although they produce worse fits to the CO SLED.
On the other hand we can reject the extended diffuse XDR emission (C model) since the observed CO surface brightness in our galaxies ($>$10$^{-6}$\,erg\,cm$^{-2}$\,s$^{-1}$\,sr$^{-1}$) is almost two order of magnitude higher than that of the best-fit model ($5\times 10^{-8}$\,erg\,cm$^{-2}$\,s$^{-1}$\,sr$^{-1}$).

\section{Summary and Conclusions}

We obtained the CO SLED (from $J_{\rm up}$ = 1 to 12) of six nearby Seyfert galaxies using SPIRE\slash FTS spectroscopic observations. The average CO SLED was fitted using simple radiative transfer models (RADEX) as well as more elaborate PDR and XDR models.

From the RADEX modeling we derived a kinetic temperature of 500$^{+500}_{-250}$\,K and H$_2$ density of 10$^{3.7\pm 1.0}$\,cm$^{-3}$. However, we note that an homogeneous model with a single temperature and gas density may not be a realistic representation of the molecular gas conditions in a Seyfert galaxy.

PDR models can explain the observed CO SLED up to $J_{\rm up}\sim 8$. Higher $J$ CO lines are underpredicted by these models and another component (PDR or XDR) is needed to reproduce the SLED. 
We also show that XDR models can fit the CO SLED, however the measured $L_{\rm CO}$\slash $L_{\rm X}$ ratios disagree with those expected for these models. Nevertheless, there are XDR models compatible with the observed $L_{\rm CO}$\slash $L_{\rm X}$ ratios, although they do not fit well the CO SLED.
In summary, an individual PDR or XDR model cannot explain the observed CO emission in these local Seyfert galaxies and, at least, two components (two PDRs or PDR$+$XDR) are needed.

Future high sensitivity and high angular resolution observations with ALMA of the mid-$J$ CO lines will help to disentangle the AGN and star-formation CO emission, as well as, to determine the excitation mechanism and heating source of the warm molecular gas.

%
%
\small  
%
\section*{Acknowledgments}   
%
This work has been funded by the Agenzia Spaziale Italiana (ASI) under contract I/005/11/0.
SPIRE has been developed by a consortium of institutes led by Cardiff Univ. (UK) and including: Univ. Lethbridge (Canada); NAOC (China); CEA, LAM (France); IFSI, Univ. Padua (Italy); IAC (Spain); Stockholm Observatory (Sweden); Imperial College London, RAL, UCL-MSSL, UKATC, Univ. Sussex (UK); and Caltech, JPL, NHSC, Univ. Colorado (USA). This development has been supported by national funding agencies: CSA (Canada); NAOC (China); CEA, CNES, CNRS (France); ASI (Italy); MCINN (Spain); SNSB (Sweden); STFC, UKSA (UK); and NASA (USA).
This research has made use of the NASA/IPAC Extragalactic Database (NED) which is operated by the Jet Propulsion Laboratory, California Institute of Technology, under contract with the National Aeronautics and Space Administration.

\def\apj{ApJ}%
\def\aap{A\&A}%
\def\apjl{ApJ}%
\def\mnras{MNRAS}%
\def\apjs{ApJS}%
\def\araa{ARA\&A}%
%

%
\end{document}